\documentclass[twocolumn,english]{revtex4-2}
\usepackage[T1]{fontenc}
\usepackage[utf8]{inputenc}
\usepackage{xcolor}
\usepackage{babel}
\usepackage{amsmath}
\usepackage{amssymb}
\usepackage{graphicx}
\PassOptionsToPackage{normalem}{ulem}
\usepackage{ulem}
\usepackage[pdfusetitle,
 bookmarks=true,bookmarksnumbered=false,bookmarksopen=false,
 breaklinks=false,pdfborder={0 0 0},pdfborderstyle={},backref=false,colorlinks=true]
 {hyperref}

\makeatletter

\providecolor{lyxadded}{rgb}{0.8,0.29,0.086}
\providecolor{lyxdeleted}{rgb}{0.86,0.2,0.18}
\DeclareRobustCommand{\mklyxadded}[1]{\textcolor{lyxadded}\bgroup#1\egroup}
\DeclareRobustCommand{\mklyxdeleted}[1]{\textcolor{lyxdeleted}\bgroup\mklyxsout{#1}\egroup}
\DeclareRobustCommand{\mklyxsout}[1]{\ifx\\#1\else\sout{#1}\fi}
\usepackage{braket}

\makeatother

\begin{document}
\title{Temperature-induced optical enhancement near a localization transition}
\author{Raul Liquito$^{1}$, Miguel Gonçalves$^{2}$,Bruno Amorim$^{1}$,
Eduardo V. Castro$^{1,3}$}
\affiliation{$^{1}$Centro de Física das Universidades do Minho e do Porto, LaPMET,
Departamento de Física e Astronomia, Faculdade de Ciências, Universidade
do Porto, Rua do Campo Alegre s/n, 4169-007 Porto, Portugal}
\affiliation{$^{2}$Princeton Center for Theoretical Science, Princeton University,
Princeton NJ 08544, USA}
\affiliation{$^{3}$Beijing Computational Science and Research Center, Beijing
100084, China}
\begin{abstract}
Quasiperiodic systems are an intermediate class of systems between
periodic crystals and disordered systems, famously exhibiting metal-insulator
transitions (MITs) even in one dimension. While their transport properties
have been studied extensively, a systematic analysis of the finite-frequency
optical conductivity near the critical point has been lacking. In
this work, we carry out a detailed study of the optical conductivity
in the paradigmatic Aubry-André model. We find that the zero-temperature
low-frequency optical signal is strongly restructured by the quasiperiodic
potential, exhibiting an optical gap that closes discontinuously as
the system approaches the MIT. Most strikingly, we uncover a mechanism
for a strong enhancement of the low-frequency finite temperature optical
conductivity at certain resonant frequencies. This enhancement stems
from the thermal activation of Pauli-blocked transitions between strongly
resonant van Hove singularities. This mechanism provides new insight
into finite-frequency transport in quasiperiodic systems and a new
pathway for manipulating optical properties near a localization transition.
Furthermore, our findings establish the optical response as a powerful,
experimentally accessible tool for probing non-trivial quasiperiodicity
effects.
\end{abstract}
\maketitle

\section{Introduction\label{sec:Introduction}}

Quantum systems with quasiperiodic modulations have been a subject
of intense study for decades \citep{kohmotoCriticalWaveFunctions1987,goldmanQuasicrystalsCrystallineApproximants1993}.
In disordered systems the absence of translational symmetry leads
to the breakdown of Bloch's band theory. In one dimension (1D), any
amount of uncorrelated disorder leads to an insulating regime characterized
by exponential localization of the eigenfunctions, a phenomena known
as Anderson localization \citep{andersonAbsenceDiffusionCertain1958,mackinnonOneParameterScalingLocalization1981}.
In contrast, translation symmetry breaking in quasiperiodic systems
(QPS) can give origin to Bloch-wave-like properties as well as Anderson
localization, famously exhibiting metal-insulator transitions (MIT)
at a finite potential strength even in one dimension \citep{aubryAnalyticityBreakingAnderson1980}.
The exotic properties of these systems, including unique topological
features \citep{krausTopologicalStatesAdiabatic2012,krausTopologicalEquivalenceFibonacci2012,verbinObservationTopologicalPhase2013,liquitoFateQuadraticBand2024,liquitoQuasiperiodicQuadrupoleInsulators2025}
and quantum transport properties \citep{prangeLongRangeResonanceAnderson1984,royFiniteTemperatureStudy2018,rooszNonequilibriumQuantumRelaxation2014,royStudyCounterintuitiveTransport2019,bhakuniNoiseinducedTransportAubryAndreHarper2024,lahiriAcConductivityIncommensurate1996,gopalakrishnanSelfdualQuasiperiodicSystems2017}
can be experimentally probed in diverse platforms, from optical \citep{roatiAndersonLocalizationNoninteracting2008,modugnoExponentialLocalizationOnedimensional2009,schreiberObservationManybodyLocalization2015,luschenObservationSlowDynamics2017,andersonConductivitySpectrumUltracold2019,kamkouSpectralDynamicalCharacters2023}
and photonic lattices \citep{lahiniObservationLocalizationTransition2009,verbinObservationTopologicalPhase2013}
to moiré materials \citep{MoireMagicThree2021,moonQuasicrystallineElectronicStates2019,yuDodecagonalBilayerGraphene2019,pezzini30degTwistedBilayerGraphene2020,goncalvesIncommensurabilityinducedSubballisticNarrowbandstates2021,uriSuperconductivityStrongInteractions2023,laiImagingSelfalignedMoire2023a}.

The Aubry-André model is the paradigmatic example describing a system
ofnon-interacting particles in a 1D lattice subjected to an on-site
potential with a quasiperiodic modulation, incommensurate with the
underlying lattice. The model hosts a MIT at a critical value of the
on-site potential strength. At weak potential strength the system
is metallic, with delocalized eigenfunctions and ballistic transport.
After the MIT, it enters an insulating/localized regime where the
eigenstates become exponentially localized, as in the Anderson model.
Exactly at the critical point, the eigenfunctions become multifractal
\citep{kohmotoCriticalWaveFunctions1987,goncalvesCriticalPhaseDualities2023,oliveiraLocalDensityStates2025}
which leads to anomalous transport \citep{purkayasthaAnomalousTransportAubryAndreHarper2018}.

Due to the exotic localization and spectral properties of quasiperiodic
systems, considerable efforts have been made towards understanding
the effects of quasiperiodicity on quantum transport. In Refs.~\citep{royStudyCounterintuitiveTransport2019,royFiniteTemperatureStudy2018}
the authors studied the persistent current by analyzing the system's
response under a magnetic flux, showing that throughout the metallic
phase the system exhibits a finite persistent current that vanishes
in the insulating regime. . Another common approach to transport
in quasiperiodic systems is by simulating diffusing wavepackets and
extracting the diffusion coefficient for distinct regimes \citep{barlevTransportQuasiperiodicInteracting2017,purkayasthaAnomalousTransportAubryAndreHarper2018,bhakuniNoiseinducedTransportAubryAndreHarper2024}
or by performing conductance calculations \citep{sutradharTransportMultifractalityBreakdown2019}.
In particular, the Aubry-André model exhibits ballistic transport
in the extended regime, is insulating in the localized regime and
superdiffusive at the critical point. All these methods primarily
probe the DC transport properties of the system. Frequency resolved
studies of the optical conductivity, on the other hand, have seen
little discussion \citep{prangeLongRangeResonanceAnderson1984,lahiriAcConductivityIncommensurate1996,gopalakrishnanSelfdualQuasiperiodicSystems2017,iijimaOpticalResponseTightbinding2022}.
Significant efforts have been made to adapt Mott's argument to quasiperiodic
systems. Mott's original argument for the 1D Anderson model \citep{mottConductionNoncrystallineMaterials1969}
relies on a homogeneous density of states (strong disorder limit)
to predict the low-frequency optical conductivity scaling $\sigma(\omega)\propto\omega^{2}\ln\left(\omega/\omega_{0}\right)^{d+1}$.

Recent works have successfully modified this approach for specific
gapless quasiperiodic system. For instance, in Ref.~\citep{prangeLongRangeResonanceAnderson1984}
the authors examine the Maryland model, an exactly solvable system
where a nearly diverging quasiperiodic potential induces a smooth,
gapless density of states and spectrum-wide Anderson localization.
Applying a Mott-like argument yields $\sigma(\omega)\propto e^{-\left(\frac{\omega_{0}}{\omega}\right)^{1/\sigma}}$.
Similarly, in Ref.~\citep{gopalakrishnanSelfdualQuasiperiodicSystems2017}
the author study a modified Aubry-André model with power-law decaying
hoppings (exponent $p\geq2$), finding $\sigma(\omega)\propto\omega^{2-3/p}$.

Extending Mott's argument to standard quasiperiodic systems is fundamentally
constrained, since most of these systems, including the Aubry-André
model, feature gapped energy spectra that violates the requirement
of a homogenous density of states. Thus, a systematic study of the
optical conductivity across the metallic-to-critical regime remains
a significant gap in the literature.

\begin{figure}[h]
\begin{centering}
\includegraphics{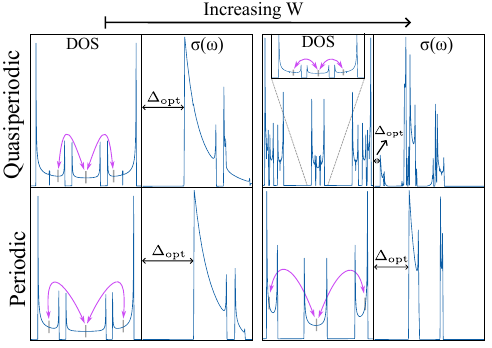}
\par\end{centering}
\caption{Schematic representation of the regular part of the real optical conductivity
as a function of frequency for increasing quasiperiodic potential
strength $W$ (left to right). The top row corresponds to the standard
Aubry--André model, the bottom row to a purely periodic harmonic
modulation with a 55-site unit cell. In the density of states panels
(DOS) the purple arrows indicate the transitions associated with the
corresponding optical gap $\Delta_{\text{opt}}$.\label{fig:Scheme_vs_W}}
\end{figure}

\begin{figure}[h]
\begin{centering}
\includegraphics{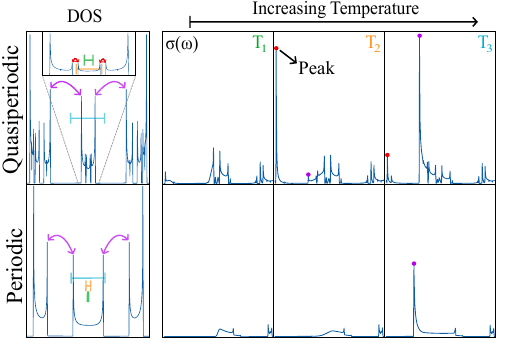}
\par\end{centering}
\caption{Schematic representation of the regular part of the real optical conductivity
as a function of frequency for increasing temperature $T$ (left to
right). The top row corresponds to the standard Aubry--André model,
the bottom row to a purely periodic harmonic modulation with unit
cell enlarged to $55$ sites. In the density of states panels (DOS)
the purple and red arrows indicate the transitions associated with
the corresponding resonance, the spacers show the width of $k_{b}T$
and are color coded to correspond to their temperatures (top right).
The frequency scale is identical across all panels, and the conductivity
scale is preserved. Curves correspond to $W=1.95$ and $T\in\{10^{-4};4\times10^{-4};3.2\times10^{-3}\}$.\label{fig:Scheme_vs_T}}
\end{figure}

In this work we address this gap by demonstrating that the unique
spectral properties of the quasiperiodic Aubry-André model give rise
to highly tunable and enhanced optical transport features not present
in periodic systems. Our primary findings are concisely illustrated
in Figs.~\ref{fig:Scheme_vs_W}~and~\ref{fig:Scheme_vs_T}, where
we schematically depict the behavior of the regular part of the absorptive
conductivity ($\text{Re}\left[\sigma_{\text{reg}}(\omega)\right]$)
near the critical point.

First, we find that the low-frequency behavior of the optical conductivity
is strongly modified as we approach the MIT from the extended phase.
As schematically illustrated in Fig.~\ref{fig:Scheme_vs_W}, this
is characterized by strong optical signals at progressively smaller
$\omega$, leading to an effectively discontinuous closing of the
optical gap $\Delta_{\text{opt}}$. This behavior is a consequence
of the successive opening of higher order spectral gaps as the critical
point is approached, ultimately leading to a fractal spectrum exactly
at the critical point. In sharp contrast, a periodic system with approximately
the same potential wavelength exhibits a robust and finite optical
gap.

Secondly, we uncovered a mechanism for a strong enhancement of the
low-frequency optical conductivity upon simply changing the temperature.
At zero temperature, transitions between adjacent, gap separated van
Hove singularities, are forbidden by Pauli exclusion. However, as
depicted in Fig.~\ref{fig:Scheme_vs_T}, a finite temperature introduces
a thermal occupation imbalance, activating these transitions. This
results in the emergence of sharp resonant peaks at finite frequencies
(purple and red dots), associated with the energy of the activated
van Hove singularities (see the temperature color encoding of Fig.~\ref{fig:Scheme_vs_T}).
This effect is strongly enhanced and far more tunable in the quasiperiodic
case compared to an periodic approximant system. It is highly sensitive
to both temperature and potential strength, which offer two simple
knobs to significantly tune the optical response.

The detailed description and in-depth analysis of the mechanism responsible
for these observations are provided in the subsequent sections.

This manuscript is organized as follows: in Sec.~\ref{sec:Aubry-Andr=0000E9-Model}
we introduce the Aubry-André (AA) model and detail numerical implementations
aspects; in Sec.~\ref{sec:Kubo-Greewood-Conductivity} we provide
a concise theoretical introduction to the Kubo-Greenwood formula used
for transport calculations; in Sec.~\ref{sec:Results} we present
our main findings and provide a detailed analysis of the AA model's
optical properties through calculations of DC and AC optical conductivity;
in Sec.~\ref{sec:Conlcusion} we summarize and discuss the main conclusions
of this work. We also include an appendix with the following sections:
in Appendix.~\ref{sec:Perturbation-Theory} we derive generic properties
of the AA eigenfunctions in the metallic regime using perturbation
theory, and obtain the perturbed current matrix elements, establishing
a connection with the Drude weight; in Appendix.~\ref{sec:Drude-Weight-at-zero-T}
we show that at zero temperature the Drude weight is directly determined
by the Fermi velocity; in \ref{sec:VHS-in-1D} we calculate the scaling
law of the number of states in a van Hove singularity with system
size for a generic one-dimensional model (important for understanding
the scaling behavior of the optical conductivity in thermal resonances
); in Appendix.~\ref{sec:Summing-Conductivity} we predict the scaling
behavior of the optical conductivity in thermal resonances; in Appendix.~\ref{sec:Temperature-Evolution-of}
we provide a detailed evolution of the optical conductivity with temperature
for two values of potential strength.

\section{Aubry-André Model\label{sec:Aubry-Andr=0000E9-Model}}

We consider the Aubry-André (AA) model threaded by a flux $\theta$,
which is described the the following Hamiltonian:

\begin{equation}
H=-t\sum_{j}\left(e^{i\frac{\theta}{L}}c^{\dagger}_{j+1}c_{j}+\text{h.c.}\right)+W\sum_{j}\cos\left(2\pi\beta j+\phi\right)c^{\dagger}_{j}c_{j}\label{eq:AA_model_Hamiltonian}
\end{equation}

where $L$ is the number of sites, $c^{\dagger}_{j}$ creates an electron
at site $j$, the first term describes hoppings between nearest-neighboring
sites, and the second term describes a quasiperiodic potential with
strength $W$ and wavenumber $2\pi\beta$. For an irrational $\beta$,
the single-particle wavefunctions are delocalized for $W<2t$ and
localized for $W>2t$. The quantum critical point ($W=2t$) is characterized
by a fractal spectrum and multifractal wave functions. The Aubry-André
model has a duality between extended and localized phases, by replacing
$W\leftrightarrow2t$, with the critical point $W=2t$, being self-dual.
Through out the manuscript, we will set $\beta=\frac{1+\sqrt{5}}{2}$,
the golden mean. .

\subsection{Spectral properties}

The Aubry-André model is the paradigmatic example that a fully deterministic
potential can also induce a metal-insulator transition. While random
disorder tends to wash out spectral features, eventually closing all
energy gaps, a quasiperiodic potential instead opens gaps on ever
finer energy scales, yielding an increasingly fragmented spectrum
that becomes a self-similar Cantor set at the critical point \citep{avilaTenMartiniProblem2009}.

\begin{figure}[h]
\begin{centering}
\includegraphics{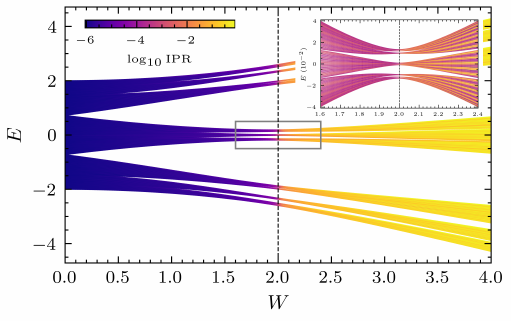}
\par\end{centering}
\caption{Evolution of AA model energy spectra as a function of quasiperiodic
potential strength $W$ for a system size $L=610$. The inset we plot
the same information for a system size $L=6765$, in the central region
highlighted by the grey rectangular box. Notice the different energy
($y$-axis) scales of the main plot and inset. The color plot encodes
the inverse participation ratio (IPR) of each eigenstate. The dashed
black vertical line marks the metal-insulator transition of the AA
model at $W=2.0$\label{fig:AA_energy_spectra}.}
\end{figure}

These complex spectral properties can be understood by considering
the potential's effect on the unperturbed Bloch states. In perturbation
theory (see Appendix~\ref{sec:Perturbation-Theory}), the potential
couples states with momenta $k$ and $k\pm2\pi m\beta$.

A simple pedagogical case is the commensurate potential with $\beta=1/2$,
which doubles the unit cell of the system. Here, lowest-order and
only coupling ($m=1$) connects $k$ and $k\pm\pi$, causing band
folding at the new Brillouin zone edge ($k=\pm\pi/2$) and opening
a gap. Increasing $W$ in this commensurate case renormalizes the
energy spectra while keeping the overall periodic structure.

\begin{figure}[h]
\begin{centering}
\includegraphics{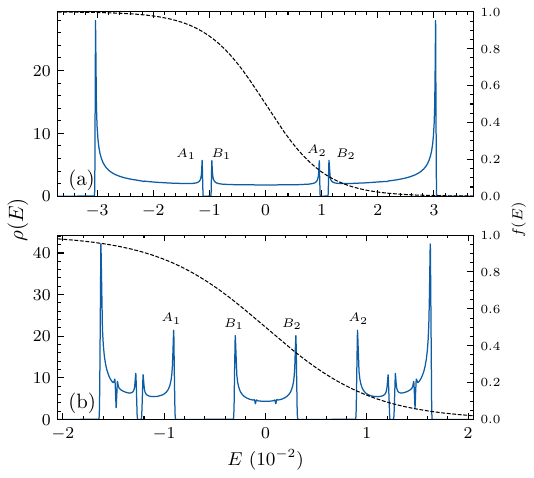}
\par\end{centering}
\caption{The plots show the DOS of the central bands around $E=0$. The dotted
black curve is the Fermi Dirac function plotted for $T\approx5\times10^{-3}$.
Some van Hove singularities are labeled to facilitate the results
analysis. The density of states were computed from the exact eigenstates
of the AA model for (a)$W=1.7$ and (b)$W=1.9$ and for a system size
of $L=28657$.\label{fig:DOS_AA}}
\end{figure}

In a truly QPS, all harmonics $m\in\mathbb{Z}$ can in principle contribute.
However, the extended/ballistic phase at $W<2t$ is characterized
by a finite correlation length and the support of the eigenfunctions
at momenta $k\pm2\pi m\beta$ decays exponentially with $m$. Because
of this, there is effectively only a finite number of harmonics contributing
and therefore a finite number of gaps being opened (all higher-order
gaps are exponentially suppressed). The resulting energy spectrum
in the ballistic regime is therefore a set continuous energy (quasi-Bloch
bands), whose dispersion at the band edges remains quadratic, giving
rise to the van Hove singularities observed in the density of states
(DOS) of the system (Fig.~\ref{fig:DOS_AA}). Exactly at the critical
point, the correlation length diverges and all harmonics contribute.
As a consequence, the eigenfunctions become delocalized in both momentum
and real-space.. For $W>2t$, the eigenstates become exponentially
localized in real space (and extended in momentum space) and the correlation
length is therefore again finite.

This rich evolution of the energy spectrum with $W$ is shown in Fig.~\ref{fig:AA_energy_spectra}.
The MIT is evident through the sharp amplitude difference in the inverse
participation ratio of the eigenstates, defined for a normalized eigenstate
$|\psi_{m}\rangle=\sum_{j}\psi^{m}_{j}|j\rangle$ as $\textrm{IPR}_{m}=\sum_{j}|\psi^{m}_{j}|^{4}$,
where $|j\rangle=c^{\dagger}_{j}|0\rangle$. The central focus of
this paper is the low-frequency optical response. We will focus at
half-filling and on a small energy window around the Fermi level ($E=0$).

\subsection{Numerical Implementation}

We carried out numerical simulations for finite systems with $L$
unit cells. To correctly model the incommensurate limit while using
periodic boundary conditions, we use rational approximants. In particular,
we set the system size $L=F_{n}$ (the $n$-th Fibonacci number) and
the potential's inverse wavelength $\beta$ to be a rational approximant
of the golden ratio ($\phi_{GR}=\frac{1+\sqrt{5}}{2}$) given by $\beta_{n}=F_{n+1}/F_{n}$.
This choice ensures that every finite system we work with contains
exactly a single unit cell.

We also work with a fixed flux $\theta=\pi/2$ and phase shift $\phi=1/2$
to break time reversal and mirror symmetry lifting all the degeneracies
in the spectrum. While such calculations can be averaged over random
ensembles of $(\theta,\phi)$, we find that our main conclusions are
robust against changes in phases.

The calculation of the optical response was carried out by evaluating
it in terms of eigenstates and eigenenergies of the system. Given
that optical conductivity is dominated by eigenenergies close to the
Fermi level, only a fraction of the total number of eigenpairs needs
to be computed. The eigenpairs close to the Fermi energy ($E=0$)
were obtained using a shift-and-invert spectral transformation alongside
a Krylov-Schur susbspace method..

\section{Kubo-Greenwood Conductivity \label{sec:Kubo-Greewood-Conductivity}}

In the non-interacting limit and at low temperatures, where the phonon
corrections to the conductivity are negligible, the optical response
under an applied electric field can be obtained through the Kubo-Greenwood
formula \citep{kuboStatisticalMechanicalTheoryIrreversible1957,greenwoodBoltzmannEquationTheory1958}.
Written in the eigenbasis of the single-particle equilibrium Hamiltonian,
the frequency-dependent optical conductivity $\sigma(\omega)$, which
is a scalar for one-dimensional systems, can be written as:

\begin{equation}
\sigma(\omega)=-\frac{i}{L}\sum_{nm}\left|j_{nm}\right|^{2}\frac{f_{n}-f_{m}}{\epsilon_{n}-\epsilon_{m}}\frac{1}{\omega+i\eta+\epsilon_{n}-\epsilon_{m}},\label{eq:Kubo_Greenwood_Conductivity}
\end{equation}

where we used units with $e=\hbar=a=1$ where $a$ is the lattice
constant, $\epsilon_{n}$ are the energy eigenvalues, $f_{n}=f(\epsilon_{n})$
the corresponding Fermi occupation probabilities, $j_{nm}=\left\langle \psi_{n}|\hat{j}|\psi_{m}\right\rangle $
the current matrix elements in the eigenbasis representation and $\eta$
the inverse mean free time / coherence time, accounting for phenomenological
dissipation. For computational purposes we choose $\eta\propto L^{-1}$
of the order of the mean level spacing at Fermi level. The conductivity
is an intensive property of a material, which can be separated in
its regular and singular part. In the limit $\eta\to0^{+}$ it takes
the form:

\begin{align}
\sigma(\omega) & =D\left[\pi\delta(\omega)+\frac{i}{\omega}\right]+\sigma_{\text{reg}}(\omega)\\
 & =\sigma_{\text{sing}}(\omega)+\sigma_{\text{reg}}(\omega),
\end{align}

where $D$ is the Drude weight (or charge stiffness). Naturally, the
Drude weight can be defined through eq.~\ref{eq:Drude_weight} as:

\begin{equation}
D=\lim_{\omega\to0}\omega\text{Im}\left[\sigma(\omega)\right].\label{eq:Drude_weight}
\end{equation}

The Drude weight is a key quantity to distinguish between metals,
dissipative metals, and insulators \citep{souzaPolarizationLocalizationInsulators2000,restaWhyAreInsulators2002,restaDrudeWeightSuperconducting2018,drudeZurElektronentheorieMetalle1900,scalapinoInsulatorMetalSuperconductor1993}.
In particular, perfect metals have finite Drude weight while insulators
have zero Drude peak.

With eq.~\ref{eq:Kubo_Greenwood_Conductivity}, and by nothing that
for degenerate states ($\epsilon_{n}=\epsilon_{m}$) the term $\frac{f(\epsilon_{n})-f(\epsilon_{m})}{\epsilon_{n}-\epsilon_{m}}$
can be replaced by the derivative $\partial f(\epsilon_{n})$, where
$\partial f(\epsilon_{n})=\left.\frac{\partial f}{\partial\epsilon}\right|_{\epsilon=\epsilon_{n}}$,
we can reach a Fermi level expression for the Drude weight given by

\begin{equation}
D=-\frac{1}{L}\sum_{\substack{nm\\
\left(\epsilon_{n}=\epsilon_{m}\right)
}
}\left|j_{nm}\right|^{2}\partial f(\epsilon_{n}),\label{eq:Drude_weight_Fermi_level}
\end{equation}

where the double sum is restrict to states with the same energy. The
singular part of the conductivity $\sigma_{\text{sing}}(\omega)$
is often called an intraband response and it is solely determined
by the Drude weight. The real part of the regular conductivity takes
the form :

\begin{equation}
\text{Re}\left[\sigma_{\text{reg}}(\omega)\right]=-\frac{1}{L}\sum_{\substack{nm\\
\left(\epsilon_{n}\neq\epsilon_{m}\right)
}
}\left|j_{nm}\right|^{2}\frac{f_{n}-f_{m}}{\epsilon_{n}-\epsilon_{m}}\frac{\eta}{\left(\omega+\epsilon_{n}-\epsilon_{m}\right)^{2}+\eta^{2}}.
\end{equation}

and accounts for interband transitions. In the results that follow,
we study the transport properties of the Aubry-André model by performing
a frequency and temperature resolved study of the Drude weight (D)
and real part of the regular conductivity.

.

\section{Results\label{sec:Results}}

We start by establishing the physical energy scales of the model.
The Hamiltonian (Eq.~\ref{eq:AA_model_Hamiltonian}) is defined by
the hopping parameter $t$, which we set as our unit of energy ($t=1$).
All other quantities, such as temperature $T$ and frequency $\omega$,
are therefore expressed in these dimensionless units. To provide a
sense of the relevant energy scales, we describe below the temperature
and frequency scales associated with typical hopping integrals of
$t\approx1\text{ eV}$:
\begin{itemize}
\item Temperature: A temperature of $T=1$ corresponds to $k_{B}T=1\text{ eV}$.
This yields $T\approx11600\text{ K}$. The finite temperatures used
in our analysis span the range $T\in[10^{-4},10^{-3}]$, corresponding
to $T\in\left[1.16,11.6\right]\,\text{K}$.
\item Frequency: A frequency of $\omega=1$ corresponds to $\hbar\omega=1\,\text{eV}$
and therefore to a frequency $\omega\approx1.52\times10^{3}\text{ THz}$.
The most relevant spectral features we will study arise for $\omega\in\left[10^{-3};10^{-2}\right]$,
corresponding to $\omega\in\left[1.52;15.2\right]\,\text{THz}$, placing
them in the far-infrared frequency range, and accessible in optical
response experimental measurements \citep{zhangTerahertzNanoimagingGraphene2018,jingTerahertzResponseMonolayer2021,guoTerahertzNanoscopyAdvances2024,chenDirectMeasurementTerahertz2024}.
\end{itemize}

\subsection{DC conductivity and Drude Weight}

We begin by analyzing the DC response, characterized by the Drude
weight, $D$. We recall that in the presence of a scattering time
$\tau$, the Drude model conductivity is given by $\sigma(\omega)=D\tau/\left(1-i\omega\tau\right)$,
such that the DC conductivity is given by $\sigma_{\text{DC}}=D\tau$.
For the metallic phase ($W<2$), the system is ballistic, and the
$T=0$ Drude weight is directly determined by the Fermi velocity,
$D=\left|v_{F}\right|/\pi$ (see Appendix.~\ref{sec:Drude-Weight-at-zero-T}
for a detailed derivation). Thus, the suppression of the DC conductivity
is equivalent to the suppression of the Fermi velocity.

\begin{figure}[h]
\begin{centering}
\includegraphics[scale=1.03]{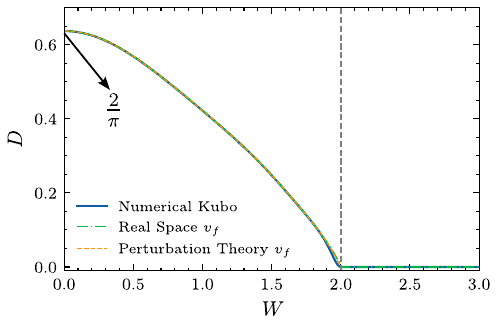}
\par\end{centering}
\caption{Drude Weight ($D$) as a function of quasiperiodic potential strength
($W$) obtained with three different methods. \textquotedblleft Numerical
Kubo\textquotedblright{} was obtained for a finite temperature of $T=5\times10^{-4}$,
a system size of $L=10946$ via Krylov-Schur sparse methods. The \textquotedblleft Real
Space $v_{F}$\textquotedblright{} was obtained by numerically calculating
the eigenstate at Fermi energy ($E=0$) for a system size $L=514229$
via via Krylov-Schur sparse methods (see Appendix.~\ref{sec:Drude-Weight-at-zero-T}).
The \textquotedblleft Perturbation Theory $v_{F}$\textquotedblright{}
curve was obtained via perturbation theory for a system size $L=1597$
truncated to order $100$ (see Appendix.~\ref{sec:Perturbation-Theory}).
The dashed grey vertical line marks the metal insulator transition
of the AA model at $W=2$. \label{fig:Drude-Weight}}
\end{figure}

In Fig.~\ref{fig:Drude-Weight}, we plot the Drude weight $D$ as
a function of $W$, evaluated using three complementary methods that
exhibit perfect agreement. The first two methods rely on the zero
temperature analytical result $D=\left|v_{F}\right|/\pi$, derived
in Appendix.~\ref{sec:Drude-Weight-at-zero-T}. In the considered
units ($t=\hbar=e=1$), the Fermi velocity is equivalent to the diagonal
current matrix element evaluated at Fermi level $J_{\epsilon_{F}}=\left\langle \psi_{\epsilon_{F}}|\hat{J}|\psi_{\epsilon_{F}}\right\rangle $.
These two approaches differ exclusively in the numerical procedure
employed to determine the Fermi-level eigenstate $\left|\psi_{\epsilon_{F}}\right\rangle $.
For the ``Real space $v_{F}$'' curve, the Hamiltonian matrix is
diagonalized in the real space basis, from which the current matrix
element is directly computed. Alternatively, the ``Perturbation Theory
$v_{F}$'' method, calculates $\left|\psi_{\epsilon_{F}}\right\rangle $
via a recursive implementation of non-degenerate, time-independent
perturbation theory (see Appendix.~\ref{subsec:Perturbation-Theory-- Numerical implementation}).
For the results presented here, the perturbative expansion is evaluated
up to the $100$th order. We emphasize that while Kohn's formula
\citep{kohnTheoryInsulatingState1964} for the Drude weight,

\begin{equation}
D\propto\left.\frac{\partial^{2}E_{0}(\theta)}{\partial\theta^{2}}\right|_{\theta=0}\label{eq:Khon_Drude_weight}
\end{equation}

requires the full many-body ground state of the system, our $T=0$
approach is formally equivalent for systems where the ground state
is a Slater determinant. This single-particle formulation offers a
significant computational advantage as it strictly requires the eigenstate
at the Fermi energy, which can be efficiently obtained utilizing sparse
Krylov-Schur eigenvalue solvers.

Lastly, the ``Numerical Kubo'' method computes the Drude weight
utilizing eq.~\ref{eq:Drude_weight_Fermi_level}. and requires a
finite temperature $T$. Because the energy derivative of the Fermi-Dirac
distribution reduces to a Dirac delta function at zero temperature,
$\partial f(\epsilon)\to-\delta(\epsilon-\epsilon_{F})$, a direct
evaluation of eq.~\ref{eq:Drude_weight_Fermi_level} at $T=0$ is
ill-posed. Thus, we evaluate the expression at a small, finite temperature
to artificially broaden the delta function. The temperature must be
sufficiently large to span several energy levels ($T>S_{\text{ave}}$,
where $S_{\text{ave}}$ is the mean level spacing around the Fermi
level) yet small enough to resolve the relevant energy scales. For
the considered system size, we set $T=5\times10^{-4}$, which is approximately
$S_{\text{ave}}$. As seen in Fig.~\ref{fig:Drude-Weight}, this
method converges to the direct $T=0$ calculations.

The monotonic decrease of $D$ in the metallic regime ($W<2$) is
a direct consequence of the effect of the quasiperiodic potential
on the band structure. As $W$ increases, the potential induces a
hierarchy of gap openings, causing a repulsion between energy bands
that in turn lead to the compression of the central band into an ever-narrower
energy window. This compression, visible in the spectral evolution
in Fig.~\ref{fig:AA_energy_spectra}, forces the quasi-Bloch bands
to become increasingly flat as $W\to2$ and, as a consequence, a smaller
$v_{F}$ and Drude weight.

This behavior is also captured perfectly by perturbation theory. As
shown in Appendix.~\ref{sec:Perturbation-Theory}, the diagonal current
matrix elements can be written as Eq.~\ref{eq:diagonal_current_matrix-elements}:

\begin{equation}
J_{k}=\left|c_{k}\right|^{2}\left(J^{(0)}_{k}+\sum^{L-1}_{n=1}\sum_{\sigma=\pm1}\left(\frac{W}{2t}\right)^{2n}\left|c_{k,\sigma,n}\right|^{2}J^{(0)}_{k+\sigma2\pi n\beta}\right),\label{eq:diagonal_current_matrix-elements}
\end{equation}

where $J^{(0)}_{k}=2\sin\left(k\right)$ are the diagonal current
matrix elements for the total current in a 1D tight binding chain,
$c_{k,\sigma,n}$ are the Bloch amplitudes of the perturbed eigenfunction
and $c_{k}$ the normalization factor (see Appendix.~\ref{sec:Perturbation-Theory}).
These are maximal at $k=\pm\pi/2$, thus near half-filling (where
$k_{F}=\pm\pi/2$). As $W$ increases, weight is transferred from
the fundamental harmonic term $J^{(0)}_{k}$ into its higher-order
harmonics $J^{(0)}_{k+\sigma2\pi n\beta}$ obeying $J^{(0)}_{k}>J^{(0)}_{k\pm2\pi n\beta}$,
for $k=\pm\pi/2$, leading to a suppression of $D$ with increasing
W .

For $W>2$, the system is in an insulating phase. All eigenstates
are exponentially localized, and the Drude weight vanishes ($D=0$).
This is a hallmark of Anderson-type localization \citep{kohnTheoryInsulatingState1964,thoulessElectronsDisorderedSystems1974},
as the diagonal current matrix elements for exponentially localized
states are suppressed .

These results agree with persistent current calculations in the AA
model \citep{royStudyCounterintuitiveTransport2019,royFiniteTemperatureStudy2018}.
In the presence of a magnetic flux $\theta$, the persistent current
can be described using a Khon-like formula derived from the Hellman-Feynaman
theorem Following eq.~\ref{eq:Khon_Drude_weight}

\begin{equation}
J_{\theta}=\left\langle \frac{\partial H_{\theta}}{\partial\theta}\right\rangle =\frac{\partial E_{0}(\theta)}{\partial\theta}\approx J_{0}+D\theta,
\end{equation}
which follows directly from eq.~\ref{eq:Khon_Drude_weight}.

\subsection{AC conductivity}

In what follows we study the real (absorptive) regular part of conductivity
($\text{Re}[\sigma_{\text{reg}}(\omega)]$) at zero and finite temperature.

\subsubsection{Zero Temperature ($T=0$)}

\begin{figure}[h]
\centering{}\includegraphics{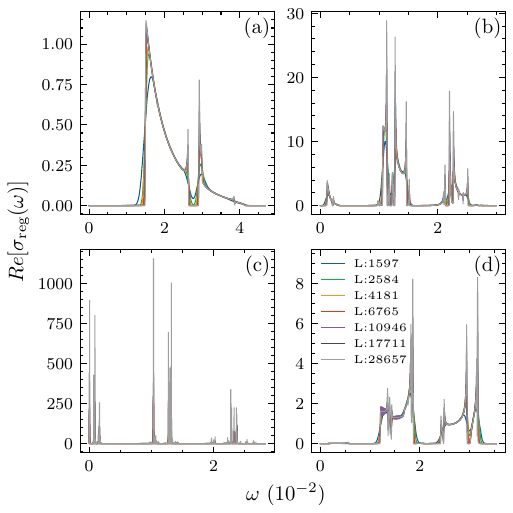}\caption{Real part of the regular conductivity ($\text{Re}\left[\sigma_{\text{reg}}(\omega)\right]$)
as a function of frequency ($\omega$) at zero temperature ($T=10^{-6}$),
for different values of potential strength. Extended phase: \textbf{(a)}
$W=1.8$; \textbf{(b)} $W=1.95$; Critical point: \textbf{(c)} $W=2.0$;
Localized phase: \textbf{(d)} $W=2.1$.\label{fig:conductivity_T=00003D0}}
\end{figure}

In the clean limit ($W=0$), the real part of $\sigma(\omega)$ exhibits
a Drude peak at $\omega\to0$ and there are no interband transitions
since there is a single energy band. In this limit, we therefore have
$\text{Re}\left[\sigma_{\text{reg}}(\omega)\right]=0$. Strictly speaking,
the current matrix elements in the eigenbasis are perfectly diagonal
and maximal for states at the Fermi energy. Increasing $W$ within
the extended phase leads to the formation of multiple bands of extended
states, which we will refer to as quasi-Bloch bands, giving rise to
finite interband optical response, as shown in Fig.~\ref{fig:conductivity_T=00003D0}(a).

Interestingly, the regular part of the conductivity exhibits a clear
optical gap ($\Delta_{\text{opt}}=1.5\times10^{-2}$ at $W=1.8$),
which is wider than the central band itself ($\approx1.2\times10^{-2}$
at $W=1.8$). This gap exists despite a finite DOS at the Fermi energy
(see Fig.~\ref{fig:DOS_AA}). The existence of a finite optical gap
throughout the metallic regime is a clear indication that quasi-Bloch
bands behave like conventional Bloch bands, and consequently all intraband
finite-frequency transitions below this gap are suppressed.

\begin{figure}[h]
\centering{}\includegraphics{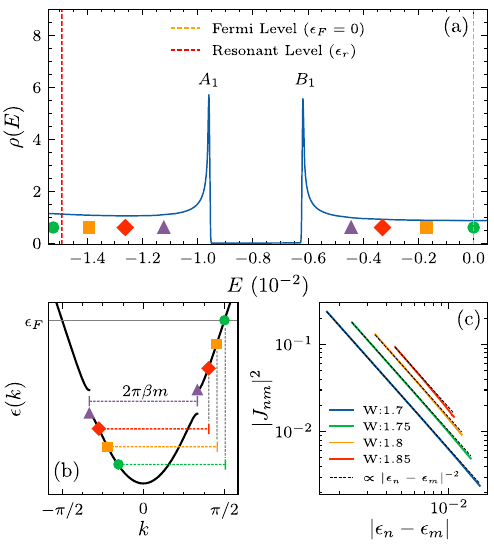}\caption{\textbf{(a)} Pairwise resonances around gapped van Hove singularities
for $W=1.8$. The solid blue line represents the DOS, the colored
markers correspond to the energy of a few eigenstates around the van
Hove singularities. Pairs with the same marker style and color indicate
strongly coupled states, i.e., with large $|J_{nm}|$. The dashed
vertical lines mark the first resonant states contributing to the
zero temperature conductivity. \textbf{(b)} Schematic representation
of the quasi-band structure around van Hove singularities and the
corresponding pairwise resonances. All the coupled states live in
distinct quasi-Bloch bands with a momentum difference of $2\pi\beta m$
for a given $m\in\mathbb{Z}$. \textbf{(c) }Off-diagonal current
matrix elements between strongly coupled resonating pairs ($|J_{nm}|^{2}$)
as a function of their energy difference ($|\epsilon_{n}-\epsilon_{m}|$)
for different values of $W$. The black dashed lines mark the expected
behavior $\left|J_{nm}\right|^{2}\propto|\epsilon_{n}-\epsilon_{m}|^{-2}$.
 \label{fig:Pairwise-resonances}}
\end{figure}

The evolution of the optical gap as a function of $W$ can be understood
through perturbation theory. The quasiperiodic potential causes states
to couple and repel, opening energy gaps and producing van Hove singularities
like $A_{1}$ and $B_{1}$ shown in Fig.~\ref{fig:DOS_AA}. This
intricate structure is governed by two primary properties:
\begin{enumerate}
\item Intraband Decoupling: states in the same quasi-Bloch band are approximatelyuncoupled.
This is analogous to an effective periodic system with a mini-Brillouin
zone (MBZ), where coupled states share the same MBZ crystalline momentum
but have different band indexes. Because the current operator is diagonal
with respect to the MBZ crystalline momenta, off-diagonal current
matrix elements vanish exponentially.
\item Interband Pairwise Resonances: States across a gap (e.g. states near
$A_{1}$ and $B_{1}$) exhibit strong pairwise resonances, leading
to off-diagonal matrix elements inversely proportional to their energy
separation ($\left|J_{nm}\right|\propto(\epsilon_{n}-\epsilon_{m})^{-2}$)\footnote{This holds for resonating pairs around gapped van Hove singularites.},
see Fig.~\ref{fig:Pairwise-resonances}(c).
\end{enumerate}

To better understand the mechanism behind the interband pairwise resonances,
consider two quasi-Bloch states with fundamental crystalline momenta
$k$ and $k'=k+2\pi\beta m$ (where $m\in\mathbb{Z}$). In the metallic
phase ($W<2$), higher order harmonics are exponentially supressed,
validating the quasi-Bloch band picture. As the system approaches
the critical point, these higher-order harmonics exceed the coupling
threshold, causing $k$ and $k'$ to effectively couple. Strong coupling
occurs when these pairs are neighbors in $k$-space ($k'=k\pm\frac{2\pi}{L}$),
driving near-divergences in the $m$-th order perturbation coefficients.
This results in level repulsion, gap opening, and the splitting of
the quasi-band. Furthermore, if $k$ and $k'$ are strongly coupled,
the adjacent states $k\pm\frac{2\pi}{L}$ and $k'\mp\frac{2\pi}{L}$
are also strongly coupled, with a perturbation coefficient inversely
proportional to the states energy difference.

This intricate pairwise coupling is illustrated in Fig.~\ref{fig:Pairwise-resonances}(a-b)
and dictates the nature of the optical gap. Because intraband transitions
within the central band are supressed, the first current generating
transition must be an interband one. At half-filling, this involves
a transition from an occupied neighboring band (e.g. state of energy
$\epsilon_{r}$ at $A_{1}$ quasi-band) to the Fermi level, leading
to an optical gap $\Delta_{\text{opt}}=\left|\epsilon_{r}-\epsilon_{F}\right|$.
Due to the symmetric nature of these resonances, $\epsilon_{r}$ is
estimated by matching the number of states between $B_{1}$ and the
Fermi level with the number of states between $A_{1}$ and $\epsilon_{r}$\_
This counting scheme provides a formal constraint for $\epsilon_{r}$::

\begin{equation}
\int^{\epsilon_{A_{1}}}_{\epsilon_{r}}\rho(\epsilon)d\epsilon=\int^{\epsilon_{F}}_{\epsilon_{B_{1}}}\rho(\epsilon)d\epsilon,\label{eq:resonant_energy_condition}
\end{equation}

where the right-hand side counts the number of states between the
Fermi level ($\epsilon_{F}$) and van Hove singularity $B_{1}$, and
the left-hand side counts the number of states between the unknown
energy $\epsilon_{r}$ and van Hove singularity $A_{1}$.

As $W$ increases towards the critical point, couplings between $k$
and $k'=k+2\pi m\beta,\,m\in\mathbb{Z}$ become relevant at progressively
higher $m$, as it is clear from the perturbation theory analysis
detailed in Appendix.~\ref{sec:Perturbation-Theory}. As a consequence,
spectral gaps are opened progressively closer to $\omega=0$ and interband
resonances become possible at correspondingly smaller $\omega$, causing
an effectively discontinuous decrease in the optical gap $\Delta_{\text{opt}}$,
as seen in Fig.~\ref{fig:conductivity_T=00003D0}(b). At the critical
point (see Fig.~\ref{fig:conductivity_T=00003D0}(c)), the optical
gap closes since the spectrum becomes fractal, with gaps opening at
all energies. This leads to enhanced conductivity, with resonances
proliferating through the whole spectrum.

In the insulating regime (see Fig.~\ref{fig:conductivity_T=00003D0}(d)),
the conductivity maintains the overall structure of its dual counterpart,
showing a similar gapped structure and optical continuum. Despite
the qualitative resemblance, the overall strength of the optical response
is smaller due to the localized nature of the eigenfunctions.

\begin{figure}[h]
\begin{centering}
\includegraphics{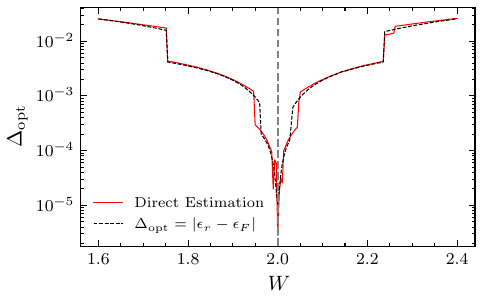}
\par\end{centering}
\caption{Optical gap ($\Delta_{\text{opt}}$) as a function of quasiperiodic
potential strength $(W)$ at zero temperature $(T=10^{-6})$. The
red line (\textquotedblleft Direct Estimation\textquotedblright )
was extracted from the conductivity calculations by imposing a threshold
for $\sigma(\omega)$ below which we took as $0$. The dotted block
line was obtained by calculating $\epsilon_{r}$ following eq.~\ref{eq:resonant_energy_condition}.\label{fig:Optical_gap}}
\end{figure}

In Fig.~\ref{fig:Optical_gap} we plot the optical gap ($\Delta_{\text{opt}}$)
as a function of $W$. To compute $\Delta_{\text{opt}}$, we consider
$\sigma\approx0$ below a cutoff of $\sigma_{\Lambda}=10^{-6}\max\left[\sigma(\omega)\right]$.
The gap computed in this way evolves through continuous regimes separated
by discontinuous jumps that become more frequent as $W\to2$, a clear
indication of the spectrum fractal nature. As $W$ increases, higher-order
terms in the perturbative expansion (see Appendix \ref{sec:Perturbation-Theory})
become relevant, effectively opening new gaps within the central band.
Each time a new gap opens, the set of allowed transitions changes,
causing $\Delta_{\text{opt}}$ to effectively jump discontinuously.
It is however important to clarify what we mean by 'effectively' here.
At any finite $W$, all orders in the perturbative expansion can in
principle contribute when the system is quasiperiodic\footnote{Note that within the localized phase ($W>2$), a large $W$ perturbative
expansion can also be carried out, treating the hopping as a perturbation.
In this case, gap openings are induced by real-space resonances (and
not momentum-space resonances like in the extended phase).}. However, away from the critical point the correlation length is
finite, so gaps generated beyond some order are exponentially suppressed.
Consequently, the optical conductivity arising from such higher-order
interband processes is nonzero but can be extremely small - below
our numerical threshold - thus producing an effective gap.

Finally, Fig.~\ref{fig:Optical_gap} also shows that the behavior
of the optical gap is similar around the self-dual point. This follows
from the duality of the AA model, which leads to identical gap opening
mechanisms in the metallic and insulating phases, with the difference
that resonances occur in momentum-space in the former, and in real-space
in the latter.

\subsubsection{Finite Temperature \label{subsec:Finite-Temperature}}

As discussed previously, states in neighboring gapped van Hove singularities
strongly couple in pairs, leading to maximum current matrix elements
between states at the edges of each quasi-Bloch band. Despite their
strong coupling, these transitions are Pauli blocked at zero temperature
and half filling.

\begin{figure}[h]
\centering{}\includegraphics{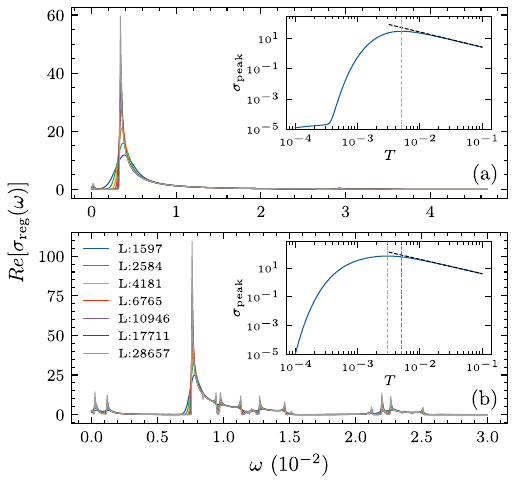}\caption{Real part of the regular conductivity ($\text{Re}\left[\sigma_{\text{reg}}(\omega)\right]$)
as a function of frequency ($\omega$) at zero temperature ($T\approx5\times10^{-3}$),
for different values of potential strength. In the insets we plot
the value of $\sigma_{\text{peak}}\equiv\max_{\omega}\text{Re}\left[\sigma_{\text{reg}}(\omega)\right]$
- the maximum value of the optical conductivity - as a function of
temperature ($T$) for the corresponding potential strength. The orange
dashed lines mark the temperature that maximizes $\sigma_{\text{peak}}$,
the gray dashed line marks $T=5\times10^{-3}$, and the black dashed
line marks the high temperature $T^{-1}$ behavior. \textbf{(a)} $W=1.8$;
\textbf{(b)} $W=1.95$.\label{fig:conductivity_finite_T}}
\end{figure}

The thermal activation of such transitions leads to prominent resonant
peaks in conductivity, which diverge in the thermodynamic limit as
seen in Fig.~\ref{fig:conductivity_finite_T}. This enhancement occurs
not only due to the strong current coupling, but also because the
states are close in energy and transitions occur between two van Hove
singularities, which contribute a large density of states. The strength
of the enhancement is controlled by both the temperature and potential
strength.

In the insets of Fig.~\ref{fig:conductivity_finite_T} we plot the
maximum of the real part of the conductivity as a function of frequency,
$\sigma_{\text{peak}}\equiv\max_{\omega}\text{Re}\left[\sigma_{\text{reg}}(\omega)\right]$,
against the temperature at fixed $W$. At very low temperature ($T\ll\omega_{\text{peak}}$),
the transitions remain effectively Pauli blocked. Increasing $T$
increases the optical conductivity until a maximum is reached at $T\approx\omega_{\text{peak}}$,
where the thermal energy ($T$) competes with the transition energy
scale. Further increasing $T$ leads to a suppression of the conductivity.
This high-temperature algebraic decay ($\sigma_{\text{peak}}\propto T^{-1}$),
marked by the dashed black line, can be obtained by expanding the
Fermi-Dirac distribution:

\begin{equation}
\frac{f_{A_{1}}-f_{B_{1}}}{\epsilon_{A_{1}}-\epsilon_{B_{1}}}\approx\left.\frac{\partial f}{\partial\epsilon}\right|_{\epsilon=\epsilon_{F}}=(4T)^{-1}.
\end{equation}

The potential strength $W$ also governs the resonance. Increasing
$W$ towards the self-dual point allows more terms to contribute to
the perturbative expansion thus increasing the current coupling between
the pairs, and leading to an increased optical response. Concurrently,
the energy spectra is compressed into narrower energy windows, lowering
the optimal resonance temperature: at $W=1.8$, the maximum resonance
occurs at $T\approx5\times10^{-3}$ as shown in Fig.~\ref{fig:conductivity_finite_T}(a),
but at $W=1.95$ it occurs at $T\approx2\times10^{-3}$, as shown
in Fig.~\ref{fig:conductivity_finite_T}(b).

\begin{figure}[h]
\centering{}\includegraphics{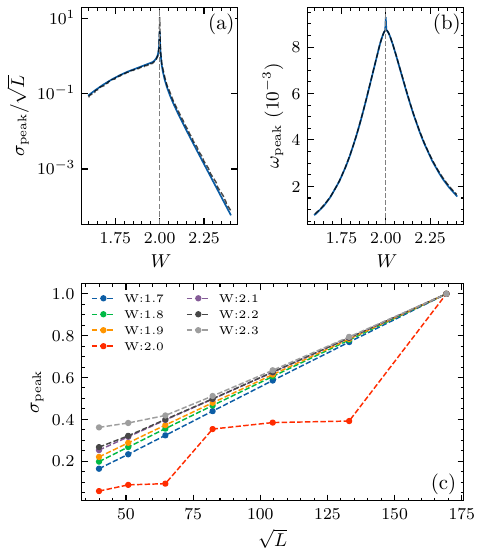}\caption{\textbf{(a)} Height of the conductivity peak ($\sigma_{\text{peak}}\equiv\max_{\omega}\text{Re}\left[\sigma_{\text{reg}}(\omega)\right]$)
as a function of potential strength ($W$) for a system with $L=28657$.
The solid line was extracted from the conductivity data, the dashed
line is a naive estimation done by explicitly summing the Kubo-Greenwood
expression for the conductivity (see Appendix.~\ref{sec:Summing-Conductivity}).
\textbf{(b)} Frequency of conductivity peak ($\omega_{\text{peak}}$)
as a function of potential strength ($W$) for a system with $L=28657$.
The solid line was extracted from the conductivity data and was obtained
by mapping the position of the resonant van Hove singularity and computing
their energy difference. \textbf{(c) }Height of the conductivity peak
($\sigma_{\text{peak}}$) as a function of $\sqrt{L}$ for different
potential strengths ($W$). The data points correspond to $L\in\{1597,2584,4181,6765,10946,17711,28657\}$.\label{fig:Peak_info}}
\end{figure}

Despite the narrowing of the quasi-Bloch bands, the resonant frequency
increases monotonically as $W$ approaches the critical point, as
seen in Fig.~\ref{fig:Peak_info}(b). This appears to contradict
the compression mechanism, however, as the energy spectra is compressed,
the increasing resonance between van Hove singularity pairs is accompanied
by an increasing repulsion, thus increasing the energy distance between
the van Hove singularities. This can be promptly observed in the DOS
in Fig.~\ref{fig:DOS_AA}.

Into the localized regime, the resonance is suppressed with increasing
$W$, as seen in Fig.~\ref{fig:Peak_info}(a). This is a clear signature
of the exponentially localized eigenfunctions. In the strong potential
limit, the localization length $\xi$ is of the order of the lattice
spacing $a$ and the current matrix elements are suppressed with increasing
potential strength. 

In Fig.~\ref{fig:Peak_info}(c), we plot the height of the peak $\sigma_{\text{peak}}$
as a function of $\sqrt{L}$ where $L$ is the tight-binding chain
size. With the exception of $W=2.0$ (self-dual point), the $\sqrt{L}$
scaling is evident indicating an intricate connection with the number
of states in each van Hove singularity, proportional to $\sqrt{L}$
(see appendix~\ref{sec:VHS-in-1D}). This stems from the pairwise
resonance structure around gapped van Hove singularities. Each state
resonates uniquely with its pair, leading to a total of number of
current generating transitions equal to the number of states in a
van Hove singularity, which is proportional to $\sqrt{L}$.It is important
to note that throughout this work we chose $\eta\propto L^{-1}$,
proportional to the mean level spacing at Fermi level, thus inducing
the $\sqrt{L}$ divergence. In fact, the resonance peaks scale as
$\sigma_{\text{peak}}\propto\eta^{-1/2}$, as shown in Appendix.~\ref{sec:Summing-Conductivity},
where is set $\eta$ is set physically by a finite coherence time
of the sample. Because of this, no true divergences are expected in
experiment.

\section{Conclusion\label{sec:Conlcusion}}

In this work, we performed a systematic study of the optical conductivity
near the metal-insulator transition of the one-dimensional Aubry-André
(AA) model.

Our analysis of the DC response demonstrated an efficient zero temperature
method to compute the Drude weight ($D$) from a single eigenstate
at the Fermi energy, enabling simulations of large-scale systems (up
to millions of

sites). This real-space approach, shown to be equivalent to perturbative
calculations of the Fermi velocity, confirms that $D$ is suppressed
as $W\to2$ due to the progressive flattening of the quasi-Bloch bands.
Our findings are in excellent agreement with previous persistent current
calculations \citep{royFiniteTemperatureStudy2018,royStudyCounterintuitiveTransport2019}.

At finite frequencies, the $T=0$ optical response serves as a powerful
probe of the model's unique spectral properties. In sharp contrast
to comparable periodic systems, we found that the optical gap $\Delta_{\text{opt}}$
is not robust under increasing potential strength. Instead, it closes
effectively discontinuously as the system approaches the critical
point, a direct consequence of the emerging fractal, Cantor-set-like
energy spectrum.

Strikingly, we observed a mechanism for strong optical enhancement
driven by finite temperature. As $T$ increases, the thermal occupation
imbalance activates previously Pauli-blocked transitions between strongly
coupled, gap-separated van Hove singularities. This results in sharp,
prominent resonant peaks in the far-infrared conductivity. This effect
is substantially more pronounced and tunable in the quasiperiodic
limit than in the periodic counterpart, providing a platform where
the resonance frequency can be precisely controlled by both potential
strength $W$ and temperature $T$.

This mechanism provides new insight into the impact of quasiperiodicity
on finite-frequency transport and offers a new pathway for manipulating
optical properties near a localization transition. While demonstrated
for the AA model, we expect this phenomenon to be a general feature
of more generic 1D quasiperiodic systems. Finally, our results identify
the optical response as a powerful experimental tool for probing quasiperiodicity,
with predictions directly testable in platforms such as ultracold
atoms in optical lattices \citep{andersonConductivitySpectrumUltracold2019,modugnoExponentialLocalizationOnedimensional2009}.
\begin{acknowledgments}
R.L. acknowledges funding from Fundação para a Ciência e Tecnologia
(FCT-Portugal) through Grant No. 2024.01276.BD. MG is supported by
a postdoctoral research fellowship at the Princeton Center for Theoretical
Science. We also acknowledge the Tianhe-2JK cluster at the Beijing
Computational Science Research Center (CSRC), the OBLIVION and VISION
supercomputer, Navigator cluster and the Deucalion cluster through
Projects No. 2022.15834.CPCA.A1 and No. 2022.15910.CPCA.A1 (based
at the High Performance Computing Center---University of \'Evora) funded
by the ENGAGE SKA Research Infrastructure (Reference No. POCI-01-0145-FEDER-022217---COMPETE
2020 and the Foundation for Science and Technology, Portugal) and
by the BigData@UE project (Reference No. ALT20-03- 0246-FEDER-000033---FEDER)
and the Alentejo 2020 Regional Operational Program.

\bibliographystyle{apsrev4-2}
\bibliography{library}

\clearpage
\onecolumngrid
\end{acknowledgments}

\appendix

\section{Perturbation Theory in the AA model \label{sec:Perturbation-Theory}}

We treat the AA model (eq.~\ref{eq:AA_model_Hamiltonian}) using
standard non-degenerate perturbation theory for $W<2t$. The unperturbed
Hamiltonian $H_{0}$ is the 1D tight-binding model, and the perturbation
is the quasiperiodic potential, $\hat{V}=W\sum_{x}\cos\left(2\pi\beta x+\phi\right)\left|x\right\rangle \left\langle x\right|$.

\subsection{Perturbed Eigenstates and Energies}

In the unperturbed Bloch basis $\left|\psi^{(0)}_{k}\right\rangle $,
the perturbation matrix element are:

\begin{equation}
\left\langle \psi^{(0)}_{k'}\left|\hat{V}\right|\psi^{(0)}_{k}\right\rangle =\frac{W}{2}\left(\delta_{k+2\pi\beta,k'}+\delta_{k-2\pi\beta,k'}\right).
\end{equation}

This shows that the perturbation only couples states with momenta
$k$ and $k\pm2\pi\beta$.

The first-order energy correction is zero: $\Delta^{(1)}_{k}=V_{kk}=0$.
The second-order correction is

\begin{equation}
\Delta^{(2)}_{k}=\sum_{k'\neq k}\frac{\left|V_{kk'}\right|^{2}}{\epsilon^{(0)}_{k}-\epsilon^{(0)}_{k'}}=\left(\frac{W}{2}\right)^{2}\sum_{\sigma=\pm1}\frac{1}{\epsilon^{(0)}_{k}-\epsilon^{(0)}_{k+\sigma2\pi\beta}}
\end{equation}

The first-order correction to the wavefunction is:

\begin{equation}
\left|\psi^{(1)}_{k}\right\rangle =\sum_{k'\neq k}\frac{V_{k'k}}{\epsilon^{(0)}_{k}-\epsilon^{(0)}_{k'}}\left|\psi^{(0)}_{k'}\right\rangle =\frac{W}{2}\sum_{\sigma=\pm1}\frac{1}{\epsilon^{(0)}_{k}-\epsilon^{(0)}_{k+\sigma2\pi\beta}}\left|\psi^{(0)}_{k+\sigma2\pi\beta}\right\rangle .
\end{equation}

The second-order correction to the wavefunction, $\left|\psi^{(2)}_{k}\right\rangle $,
couples to states $\left|\psi^{(0)}_{k\pm4\pi\beta}\right\rangle $.
All odd-order energy correction are zero, as they require returning
to the original $k$-state, which involves an even number of potential
couplings.

This process generates a perturbative expansion where the $n$-th
order wavefunction correction, $\left|\psi^{(n)}_{k}\right\rangle $,
couples the unperturbed state $\left|\psi^{(0)}_{k}\right\rangle $
to its $n$-th order harmonics, $\left|\psi^{(0)}_{k\pm2\pi n\beta}\right\rangle $,
as well as lower-order harmonics via more complex $k$-paths.

The full (unnormalized) quasi-Bloch state can be written structurally
as:

\begin{equation}
\left|\psi_{k}\right\rangle \propto\left|\psi^{(0)}_{k}\right\rangle +\sum^{+\infty}_{n=1}\left(\frac{W}{2}\right)^{n}\sum_{\sigma=\pm1}\left(c^{(n)}_{k,\sigma,n}\left|\psi_{k+\sigma2\pi n\beta}\right\rangle +\sum_{m<n}c^{(n)}_{k,\sigma,m}\left|\psi_{k+\sigma2\pi m\beta}\right\rangle \right),
\end{equation}
where the coefficients $c^{(n)}_{k,\sigma,n}=\prod^{n}_{m=1}\frac{1}{\epsilon^{(0)}_{k}-\epsilon^{(0)}_{k+\sigma2\pi m\beta}}$
come from trivial $k$-space paths, and $c^{(n)}_{k,\sigma,m}$ measure
the coupling to lower-order harmonics. Thus the coefficients $c^{(n)}_{k,\sigma,m}$
represent the $n$-th correction to the $m$-th order harmonic. From
$\Delta^{(n)}_{k}=\langle\psi^{(0)}_{k}|\hat{V}|\psi^{(n-1)}_{k}\rangle$,
the full energy correction is:

\begin{equation}
\epsilon_{k}\approx\epsilon^{(0)}_{k}+\sum_{n}\sum_{\sigma=\pm1}\left(\frac{W}{2}\right)^{2n}c^{(2n-1)}_{k,\sigma,1}.
\end{equation}

The full unnormalized quasi-Bloch state can be rewritten as a plane
wave expansion by grouping all $m$-th order corrections into a single
coefficient:

\begin{align}
\left|\psi_{k}\right\rangle  & \propto\left|\psi^{(0)}_{k}\right\rangle +\sum_{n}\sum_{\sigma=\pm1}\left(\frac{W}{2}\right)^{n}c_{k,\sigma,n}\left|\psi^{(0)}_{k+\sigma2\pi n\beta}\right\rangle \\
c_{k,\sigma,n} & =c^{(n)}_{k,\sigma,n}+\sum_{m>n}\left(\frac{W}{2}\right)^{m-n}c^{(m)}_{k,\sigma,n}
\end{align}

The normalized quasi-Bloch state is then:

\begin{equation}
\left|\psi_{k}\right\rangle =c_{k}\left|\psi^{(0)}_{k}\right\rangle +\sum_{n}\sum_{\sigma=\pm1}\left(\frac{W}{2}\right)^{n}\tilde{c}_{k,\sigma,n}\left|\psi^{(0)}_{k+\sigma2\pi n\beta}\right\rangle \label{eq:perturbation_theory_quasi_Bloch_states}
\end{equation}
where $c_{k}=\left(1+\sum_{n}\left(\frac{W}{2}\right)^{2n}\left|c_{k,\sigma,n}\right|^{2}\right)^{-1/2}$
is the normalization factor (the weight of the original Bloch state)
and $\tilde{c}_{k,\sigma,n}=c_{k}c_{k,\sigma,n}$ are the normalized
coefficients for the harmonics.

\subsection{Current Matrix Elements of quasi-Bloch states}

The current operator in the unperturbed Bloch basis is diagonal: $J^{(0)}_{kk'}=\left\langle \psi^{(0)}_{k}\left|\hat{J}\right|\psi^{(0)}_{k}\right\rangle =\delta_{kk'}J^{(0)}_{k}$,
where $J^{(0)}_{k}=\partial_{k}\epsilon_{k}=2\sin\left(k\right)$
(for $t=e=\hbar=a=1$). From Eq.~\ref{eq:perturbation_theory_quasi_Bloch_states}
the perturbed current matrix elements are:

\begin{multline}
J_{kk'}=\left|c_{k}\right|^{2}J^{(0)}_{k,k'}+\sum_{n}\left(\frac{W}{2t}\right)^{n}\left[\sum_{\sigma=\pm1}\left(c_{k'}\tilde{c}_{k\sigma m}J^{(0)}_{k+\sigma2\pi n\beta,k'}+\tilde{c}_{k'\sigma n}c^{*}_{k}J^{(0)}_{k,k'+\sigma2\pi n\beta}\right)\right]+\\
+\sum_{n,m}\left(\frac{W}{2t}\right)^{n+m}\sum_{\sigma',\sigma=\pm1}\tilde{c}_{k\sigma n}\tilde{c}_{k'\sigma'm}J^{(0)}_{k+\sigma2\pi n\beta,k'+\sigma'2\pi m\beta}.
\end{multline}

The perturbed current matrix elements have three distinct terms. The
first, represents current coupling between the clean Bloch basis and
are weighted by the normalization factor $c_{k}$. The second (middle),
are off-diagonal terms representing the current coupling between the
original Bloch state of $\left|\psi_{k}\right\rangle $ with the higher-order
harmonics of $\left|\psi_{k'}\right\rangle $ (and vice-versa). The
last term represents current coupling between the higher-order harmonics
of both perturbed states.

\subsection{Drude Weight from Perturbed Current Matrix Elements}

To obtain the Drude weight we need to calculate the Fermi velocity
Eq.~\ref{eq:Drude_weight_Fermi_velocity}. So, we are interested
in the diagonal matrix element $J_{k}=\left\langle \psi_{k}\left|\hat{J}\right|\psi_{k}\right\rangle $
for the perturbed state. Due to the translation symmetry, the unperturbed
current matrix elements are diagonal $J^{(0)}_{kk'}\propto\delta_{kk'}$
and the cross-terms in $J_{kk'}$ vanish. The resulting expression
is a weighted sum of the unperturbed current matrix elements:

\begin{equation}
J_{k}=\left|c_{k}\right|^{2}J^{(0)}_{k}+\sum_{n}\sum_{\sigma=\pm1}\left(\frac{W}{2t}\right)^{2n}\left|\tilde{c}_{k\sigma n}\right|^{2}J^{(0)}_{k+\sigma2\pi n\beta}.
\end{equation}

This is the same expression as Eq.~\ref{eq:diagonal_current_matrix-elements}
in the main text. This result directly explains the suppression of
the Drude weight $D$. At $T=0$, $D\propto\left|v_{F}\right|\propto J_{k_{F}}$
(Appendix.~\ref{sec:Drude-Weight-at-zero-T}).
\begin{enumerate}
\item For $W=0$, $c_{k}=1$ and all $\tilde{c}=0$. $J_{k}=J^{(0)}_{k}$,
which is maximal at $k_{F}=\pm\pi/2$.
\item As $W$ increases, the normalization factor $\left|c_{k}\right|^{2}$
(the weight of the original state) decreases as spectral weight is
transferred to the harmonics ($\left|\tilde{c}_{k,\sigma,n}\right|^{2}>0$).
\item The harmonics $k_{F}\pm2\pi n\beta$ are further from the band center,
so their current matrix elements $\left|J^{(0)}_{k\pm2\pi n\beta}\right|$
are smaller than $\left|J^{(0)}_{k_{F}}\right|$.
\end{enumerate}
The total current $J_{k}$ is thus a sum where weight is transferred
from the maximal term $J^{(0)}_{k_{F}}$ to progressively smaller
terms. This leads to a monotonic supression of $J_{k_{F}}$, and therefore
$D$, as $W\to2$.

\subsection{Perturbation Theory - Numerical Implementation \label{subsec:Perturbation-Theory-- Numerical implementation}}

In this section, we provide the expressions used for the recursive
implementation of time-independent non-degenerate perturbation theory
\citep{sakuraiModernQuantumMechanics2021}. The recursive implementation
allows the calculation of the perturbed eigenfunctions up to order
$m$. We used:

\begin{align}
\Delta^{(N)}_{n} & =\langle\psi^{(0)}_{n}|V|\psi^{(N-1)}_{n}\rangle\\
|\psi^{(N)}_{n}\rangle & =\sum_{i\neq n}\frac{\langle\psi^{(0)}_{i}|V|\psi^{(N-1)}_{n}\rangle}{E^{(0)}_{n}-E^{(0)}_{i}}\left|\psi^{(0)}_{i}\right\rangle -\sum_{i\neq n}\sum_{m\geq0}\Delta^{(m+1)}_{n}\frac{\langle\psi^{(0)}_{i}|\psi^{(N-1-m)}_{n}\rangle}{E^{(0)}_{n}-E^{(0)}_{i}}\left|\psi^{(0)}_{i}\right\rangle 
\end{align}

where $\Delta^{(N)}_{n}$ is the $m$-th order correction to the eigenenergy
$E_{n}=E^{(0)}_{n}+\sum_{m}\Delta^{(m)}_{n}$ and $|\psi^{(N)}_{n}\rangle$
the $m$-th order correction to the wavefunction $|\psi_{n}\rangle=|\psi^{(0)}_{n}\rangle+\sum_{m}|\psi^{(m)}_{n}\rangle$.

\section{Drude Weight at $T=0$\label{sec:Drude-Weight-at-zero-T}}

In Sec.~\ref{sec:Kubo-Greewood-Conductivity} the Drude weight was
introduced. For a translationally invariant system, the Hamiltonian
is diagonal in the Bloch basis, and Eq.~\ref{eq:Drude_weight_Fermi_level}
takes the form:

\begin{equation}
D=-\frac{1}{L}\sum_{k\in\text{FBZ}}\frac{\partial f}{\partial\epsilon_{k}}\left|J_{k}\right|^{2}.
\end{equation}

In the thermodynamic limit $(L\to\infty)$ the sum becomes an integral:

\begin{equation}
D\approx-\frac{1}{2\pi}\int_{\text{FBZ}}dk\frac{\partial f}{\partial\epsilon_{k}}\left|J_{k}\right|^{2}.
\end{equation}

We can take the zero temperature limit by noting that the Fermi-Dirac
derivative becomes a delta function: $-\frac{\partial f}{\partial\epsilon_{k}}\to\delta\left(\epsilon_{k}-\epsilon_{F}\right)$.
At half-filling ($\epsilon_{F}=0$ for the clean system), the integral
is:

\begin{equation}
D=\frac{1}{2\pi}\int_{\text{FBZ}}\delta\left(\epsilon_{k}-\epsilon_{F}\right)\left|J_{k}\right|^{2}dk.\label{eq: thermo_limit_drude_wieght}
\end{equation}

To solve this integral, we use the identity $\delta\left(g(k)\right)=\sum_{i}\frac{\delta\left(k-k_{i}\right)}{\left|g'(k_{i})\right|}$,
where $k_{i}$ are the roots of $g(k)$. Here, $g(k)=\epsilon_{k}-\epsilon_{F}$,
and the roots are the Fermi momenta, $k_{F}=\pm\pi/2$. The derivative
is $g'(k)=\partial_{k}\epsilon_{k}=v_{k}$ (the group velocity). At
the Fermi points, $\left|v\left(\pm\pi/2\right)\right|=\left|v_{F}\right|$.
The delta function is therefore:

\begin{equation}
\delta\left(\epsilon_{k}-\epsilon_{F}\right)=\frac{\delta\left(k-\pi/2\right)}{\left|v_{F}\right|}+\frac{\delta\left(k+\pi/2\right)}{\left|v_{F}\right|}.
\end{equation}

Substituting this into Eq.~\ref{eq: thermo_limit_drude_wieght} yields:

\begin{align}
D & =\frac{1}{2\pi\left|v_{F}\right|}\int_{\text{FBZ}}\left[\delta\left(k-\pi/2\right)+\delta\left(k+\pi/2\right)\right]\left|J_{k}\right|^{2}dk\\
 & =\frac{1}{2\pi\left|v_{F}\right|}\left(\left|J_{-\pi/2}\right|^{2}+\left|J_{\pi/2}\right|^{2}\right).
\end{align}

Since $\left|J_{-\pi/2}\right|=\left|J_{\pi/2}\right|$, the two Fermi
points contribute equally:

\begin{equation}
D=\frac{1}{\pi}\frac{\left|J_{k_{F}}\right|^{2}}{\left|v_{F}\right|}.\label{eq:thermo_drude_wieght_mid1}
\end{equation}

In units of $e=\hbar=a=1$, the current operator is $J_{k}=\partial_{k}\epsilon_{k}=v_{k}$.
Therefore, $J_{k_{F}}=v_{F}$. Substituting into Eq.~\ref{eq:thermo_drude_wieght_mid1},
gives the final well-known result:

\begin{equation}
D=\frac{\left|v_{F}\right|}{\pi}\label{eq:Drude_weight_Fermi_velocity}
\end{equation}

This result is valid for any 1D system at $T=0$. For the clean 1D
tight-binding chain, the dispersion relation is $\epsilon_{k}=-2t\cos\left(k\right)$,
and so we have $\left|v_{F}\right|=\left|v\left(k_{F}=\pm\pi/2\right)\right|=2t$.
With $t=1,$this yields the maximum Drude weight $D=\frac{2}{\pi}\approx0.637$.

For the AA model with $W<2$, the system remains ballistic, and the
unperturbed momentum $k$ remains a good quantum number. Therefore,
Eq.~\ref{eq:Drude_weight_Fermi_velocity} still holds, however, with
a $W$ dependent Fermi velocity $v_{F}$. As described in Sec.~\ref{sec:Drude-Weight-at-zero-T},
this $v_{F}$ can be obtained either from perturbation theory or from
real-space numerical calculations of $J_{k_{F}}=\left\langle \psi_{\epsilon_{F}}\left|\hat{J}\right|\psi_{\epsilon_{F}}\right\rangle $.

\section{van Hove singularities in 1D\label{sec:VHS-in-1D}}

We derive the density of states (DOS) for a 1D tight-binding chain
and calculate the scaling of the number of states near a van Hove
singularity. The density of states is given by:

\begin{align}
\rho(E) & =\frac{1}{L}\sum_{k\in FBZ}\delta\left(E-\epsilon(k)\right)\\
 & =\frac{1}{2\pi}\int^{\pi}_{-\pi}\delta\left(E-\epsilon(k)\right)dk.\label{eq:DOS_k_space}
\end{align}

For the 1D chain, $\epsilon(k)=-2tcos(k)$. Using the identity $\delta$$\left(g(k)\right)=\sum_{i}\frac{\delta(k-k_{i})}{\left|g'(k_{i})\right|}$,
and $\partial_{k}\epsilon_{k}=2t\sin\left(k\right)$ leads to the
well-known result:

\begin{align}
\rho(E) & =\frac{1}{\pi}\int^{\pi}_{0}\delta\left(E+2t\cos\left(k\right)\right)dk\\
 & =\frac{1}{2\pi t\sin\left(\arccos\left(-\frac{E}{2t}\right)\right)}.\\
 & =\frac{1}{\pi\sqrt{\left(2t\right)^{2}-E^{2}}}
\end{align}

This expression exhibits a square root divergence at the band edges
$E=\pm2t$, corresponding to the 1D van Hove singularities. Close
to the van Hove singularity, we have $\rho(E)\approx\frac{1}{2\pi\sqrt{t}\sqrt{2t-\left|E\right|}}$.

We can now determine how the number of states $N_{\text{vHs}}$ in
an energy window $\Delta E$ near the van Hove singularity scales
with system size $L$. We are interested in an energy window $\Delta E$
that shrinks with $L$, specifically $\Delta E\propto\left\langle S_{ave}\right\rangle \propto\frac{1}{L}$,
where $S_{ave}$ is the mean level spacing around the Fermi level.

To inspect how the number of states in the singularity changes with
system size, we evaluate the following integral,

\begin{equation}
N_{\text{VHS}}=L\int^{2t}_{2t-\Delta E}\rho(E)dE.
\end{equation}

Using the approximation $\rho(E)\approx C\left(2t-\left|E\right|\right)^{-1/2}$
(where $C$ is a constant), we solve the integral:

\begin{align}
N_{\text{VHS}} & \approx L\cdot C\int^{2t}_{2t-\Delta E}\left(2t-\left|E\right|\right)^{-1/2}dE\\
N_{\text{VHS}} & \propto L\cdot\left[-2\sqrt{2t-\left|E\right|}\right]^{2t}_{2t-\Delta E}\propto L\sqrt{\Delta E}.
\end{align}

By substituting the scaling of the energy window, $\Delta E\propto L^{-1}$,
we obtain the final scaling for the number of states within the singularity:

\begin{equation}
N_{\text{VHS}}\propto L\sqrt{1/L}\propto\sqrt{L}.
\end{equation}

This $\sqrt{L}$ scaling is a direct result of the quadratic band
edge, and is used in Appendix~\ref{sec:Summing-Conductivity} to
determine the scaling of the conductivity peak.

\section{Conductivity Peak Scaling\label{sec:Summing-Conductivity}}

At finite temperatures, transitions between adjacent, partially filled
van Hove singularities lead to strong resonant peaks in the conductivity
. To understand the system-size scaling of these peaks (Fig.~\ref{fig:Peak_info}(c)),
we provide a heuristic derivation based on the Kubo-Greenwood formula.
In the Kubo-Greenwood expression (Eq.~\ref{eq:Kubo_Greenwood_Conductivity}),
the phenomenological broadening $\eta$ represents an energy scale
below which discrete energy levels are mixed. For a finite system
of size $L$, the mean level spacing $\left\langle S_{\text{ave}}\right\rangle $
scales as $\left\langle S_{\text{ave}}\right\rangle \propto L^{-1}$.
To properly capture the thermodynamic limit, $\eta$ must be related
to this intrinsic energy scale. We thus set the broadening proportional
to the mean level spacing: $\eta\propto L^{-1}$.

We simplify our problem by considering only the two resonating van
Hove singularities, $A$ and $B$, separated by $\omega_{\text{peak}}$.
As shown in Appendix.~\ref{sec:VHS-in-1D}, the number of states
$N_{\text{vHs}}$ within a small energy window $\Delta E$ from a
1D van Hove singularity diverges as $N_{\text{vHs}}\propto L\sqrt{\Delta E}$.
We can approximate the sum in the Kubo formula (Eq.~\ref{eq:Kubo_Greenwood_Conductivity})
at the resonance $\omega=\omega_{\text{peak}}$by an average over
the current-generating transitions:

\begin{equation}
\sigma_{\text{peak}}=\text{Re}\left[\sigma_{\text{reg}}(\omega_{\text{peak}})\right]\approx\frac{\left(f_{1}-f_{2}\right)\left|\left\langle J_{\text{ave}}\right\rangle \right|^{2}N_{J_{\neq0}}}{\eta\omega_{peak}L},\label{eq:sum_kubo_cond_VHs}
\end{equation}

where $N_{J\neq0}$ is the total number of current-generating transitions.
As argued in the main text, states in adjacent van Hove singularities
are coupled in strong pairwise resonances. Therefore, $N_{J\neq0}\neq N^{2}_{\text{VHS}}$,
but instead it is proportional to the number of states in a single
van Hove singularity, $N_{J\neq0}\propto N_{\text{vHs}}$. The relevant
energy window for these transitions is the broadening $\eta$. Thus,
the number of contributing states is $N_{J\neq0}\propto L\sqrt{\Delta E}$.
Substituting this into Eq.~\ref{eq:sum_kubo_cond_VHs}, we find the
scaling of the peak:

\begin{equation}
\sigma_{\text{peak}}\propto\frac{1}{L}\frac{\left|\left\langle J_{ave}\right\rangle \right|^{2}L\sqrt{\eta}}{\eta}\propto\eta^{-1/2}.
\end{equation}

For a generic algebraic scaling $\eta\propto L^{-\alpha}$ with $\alpha\geq0$
we get $\sigma_{\text{peak}}\propto L^{\alpha/2}$ which agrees with
our $\sqrt{L}$ scaling if we choose $\eta\propto\left\langle S_{\text{ave}}\right\rangle \implies\alpha=1$.
This result, derived from the unique pairwise coupling and van Hove
singularities structure of the 1D system, is in excellent agreement
with the numerical results presented in Fig.~\ref{fig:Peak_info}(c).

\section{Temperature Evolution of Conductivity\label{sec:Temperature-Evolution-of}}

To supplement the analysis of Sec.~\ref{subsec:Finite-Temperature},
we provide the detailed evolution of the real part of the regular
conductivity, $\text{Re}\left[\sigma_{\text{reg}}(\omega)\right]$,
as a function of increasing temperature for two representative potential
strengths.

\begin{figure}[h]
\begin{centering}
\includegraphics{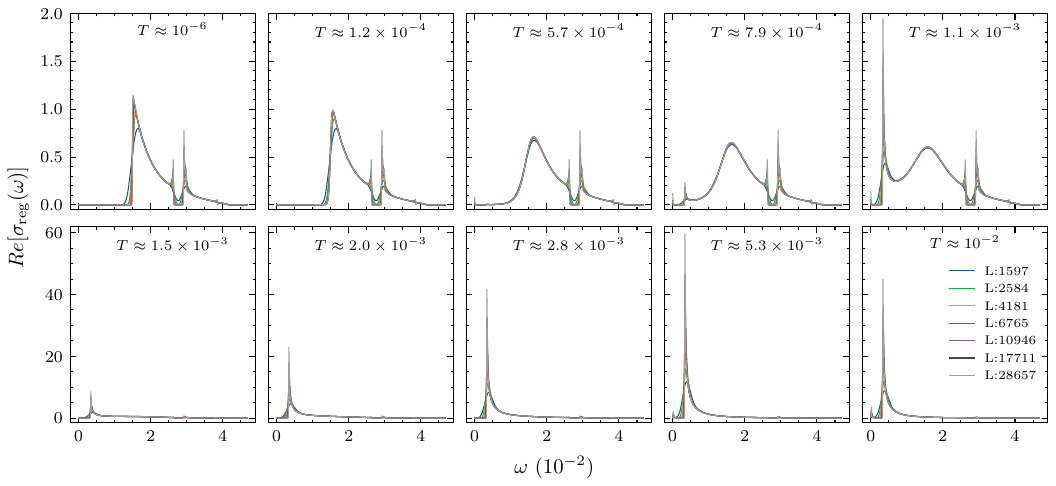}
\par\end{centering}
\caption{Real part of the regular conductivity ($\text{Re}\left[\sigma_{\text{reg}}(\omega)\right]$)
as a function of frequency ($\omega$) at $W\approx1.8$ for different
values of increasing temperature. The top and bottom row of plots
are on different scales (see $y$-axis) but share the same scale row
wise. The $x$-axis is shared between all plots. \label{fig:Cond_vs_freq_squential_temperature_W1.8}}
\end{figure}

For $W\approx1.8$ (see Fig.\ref{fig:Cond_vs_freq_squential_temperature_W1.8})
increasing the temperature initially smoothes the $T=0$ spectral
features. As $T$ becomes comparable to the relevant energy gaps,
strong thermal mixing occurs between states in adjacent, gapped van
Hove singularities. This leads to the emergence of a dominant resonant
peak, which reaches its maximal intensity at $T\approx5\times10^{-3}$,
roughly $60$ times larger than its zero-temperature counterpart.

\begin{figure}[h]
\begin{centering}
\includegraphics{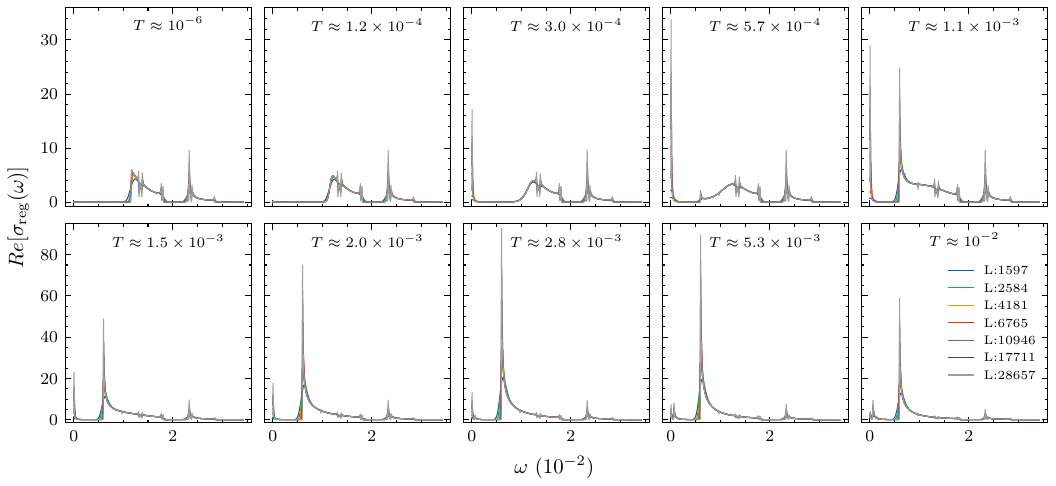}
\par\end{centering}
\caption{Real part of the regular conductivity ($\text{Re}\left[\sigma_{\text{reg}}(\omega)\right]$)
as a function of frequency ($\omega$) at $W\approx1.9$ for different
values of increasing temperature. The top and bottom row of plots
are on different scales (see $y$-axis) but share the same scale row
wise. The $x$-axis is shared between all plots.\label{fig:Cond_vs_freq_squential_temperature_W1.9}}
\end{figure}

For $W\approx1.9$ (see Fig.~\ref{fig:Cond_vs_freq_squential_temperature_W1.9}),
the conductivity exhibits sharper features at low temperature, reflecting
the increased fractal nature of the energy spectrum closer to the
critical point. As temperature increases, a similar thermal response
appears, but its maximum occurs at a lower temperature ($T\approx3\times10^{-3}$)
due to the compression of the spectral bands. Furthermore, the opening
of smaller, higher-order mid-band gaps creates a hierarchy of resonances.
A secondary, lower-frequency peak is activated at $T\approx5\times10^{-4}$.
This demonstrates a mechanism for frequency and temperature selective
optical transport.
\end{document}